\documentclass[prl,aps,twocolumn,superscriptaddress]{revtex4}
\usepackage[latin1]{inputenc} \usepackage{amsmath}
\usepackage{amsfonts} \usepackage{amssymb} \usepackage{graphicx}
\usepackage{color} \usepackage{hyperref} \usepackage{xspace}
\usepackage{dcolumn}
\usepackage{booktabs}  

\newcommand{\dd}{{\rm d}} 

\newcommand{\UNIT}[1]{\ensuremath{\,{\rm #1}}\xspace}

 \newcommand{\GeV}{\UNIT{GeV}}
 
\newcommand{\GeVfmt}{\UNIT{GeV/fm^3}} \newcommand{\TeV}{\UNIT{TeV}}

\newcommand{\fmc}{\ensuremath{\,{\rm fm}/c}\xspace}
\newcommand{\fmt}{\UNIT{fm^3}} 

\definecolor{magenta}{cmyk}{0,1,0,0}

\begin{document}

\title{Fast Dynamical Evolution of Hadron Resonance Gas via Hagedorn States}

\author{M.~Beitel} \author{C.~Greiner}  
\affiliation{Institut f\"ur Theoretische Physik, Goethe-Universit\"at
  Frankfurt am Main, Max-von-Laue-Str.~1, D-60438 Frankfurt am Main,
  Germany}
\author{H.~Stoecker}
\affiliation{Institut f\"ur Theoretische Physik, Goethe-Universit\"at
  Frankfurt am Main, Max-von-Laue-Str.~1, D-60438 Frankfurt am Main,
  Germany}
\affiliation{GSI Helmholtzzentrum f\"ur Schwerionenforschung GmbH, Planckstra\ss e  1,  D-64291
Darmstadt,  Germany }
\affiliation{Frankfurt  Institute  for  Advanced  Studies,
Ruth-Moufang-Str.    1,
D-60438  Frankfurt  am  Main,  Germany}

\begin{abstract}
Hagedorn states are the key to understand how all hadrons observed in high energy heavy ion
collisions seem to reach thermal equilibrium
so quickly. An assembly of Hagedorn states is formed 
in elementary hadronic or heavy ion collisions at hadronization. Microscopic simulations within
the transport model UrQMD allow to study the time evolution of such a pure non-equilibrated Hagedorn state
gas towards a thermally equilibrated Hadron Resonance Gas by using dynamics, which unlike strings,
fully respect detailed balance. Propagation, repopulation,
rescatterings and decays of Hagedorn states provide the yields of all
hadrons up to a mass of $m=2.5\GeV$. Ratios of feed down corrected hadron multiplicities are 
compared to corresponding experimental data from the ALICE collaboration at LHC. The quick
thermalization within $t=1-2\fmc$ of the
emerging Hadron Resonance Gas exposes Hagedorn states as a tool to understand
hadronization.  
\end{abstract}

\maketitle
\begin{figure}
\includegraphics[angle=270,width=0.50\textwidth]{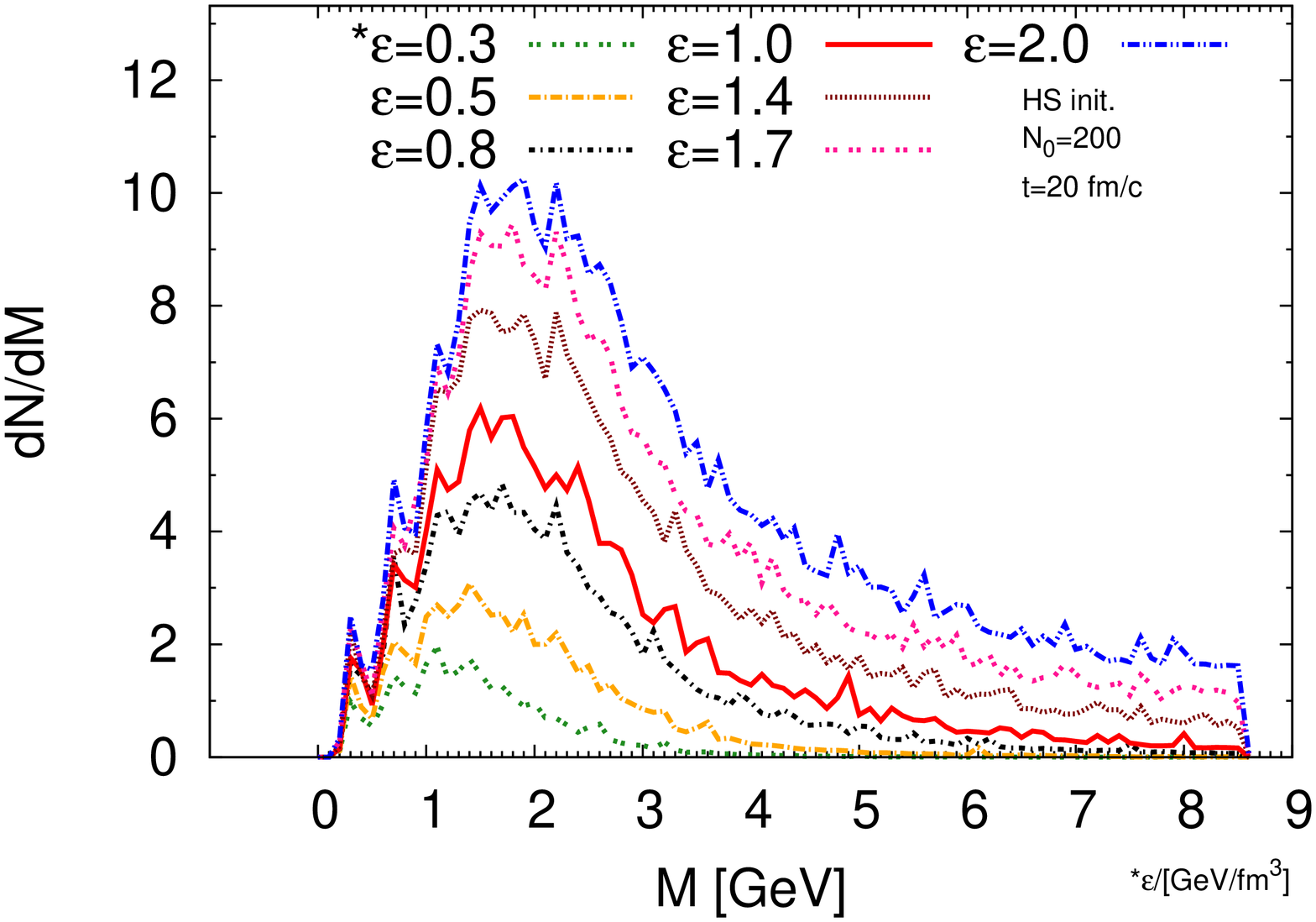}
\caption{Mass distribution of HS in thermal equilibrium $\left(t=20\fmc\right)$ for energy 
	densities in the range
$\epsilon=0.3-2.0\GeVfmt$.}
\label{fig:hsmassdistprl}
\end{figure}
\begin{figure}
\includegraphics[angle=270,width=0.50\textwidth]{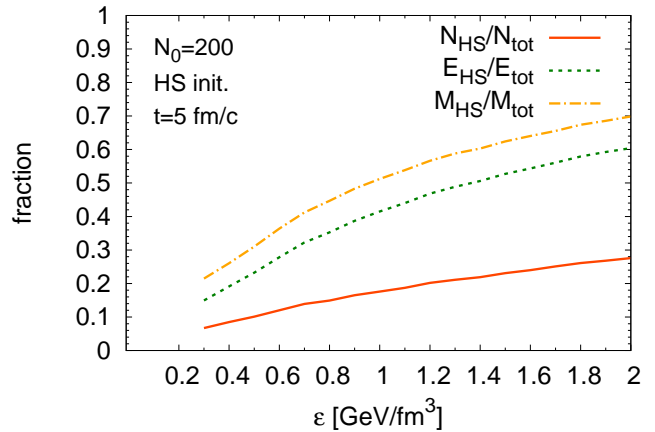}\\
\includegraphics[angle=270,width=0.50\textwidth]{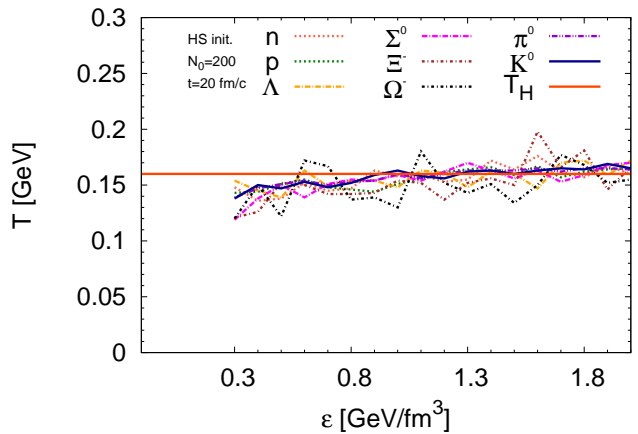}
\caption{Fraction (top) of total multiplicities, energy and mass occupied by HS in thermal
	equilibrium $\left(t\ge5\fmc\right)$ as function of energy density. Boltzmann slopes
	(temperatures) of hadrons (bottom) as function of energy density in thermal
equilibrium $\left( t=20\fmc \right)$. Red solid line denotes the Hagedorn temperature $T_H$.}
\label{fig:tempeps}
\end{figure}

Before the emergence of QCD as the theory of strong interactions physicists already developed
several models and ideas to describe observables in connection with high energetic particle
collisions of several types. A prominent phenomenological model is the Statistical Bootstrap Model
(SBM) emerging from first applications of statistical means 
\cite{Hagedorn:1965st,Hagedorn:1967ua,Hagedorn:1968jf}. Especially in (ultra-) relativistic heavy ion
collisions thermal models \cite{BraunMunzinger:2001ip,Andronic:2011yq} seem to show an excellent description 
of various hadronic particle multiplicities by choosing a temperature, volume and chemical
potentials. In this paper we provide a 
microscopic explanation for the validity of the thermal model and the very fast equilibration at
hadronization. In our approach the system temperature equals the Hagedorn temperature.
\newline
The Hagedorn temperature $T_H$ is the
highest temperature that systems of hadronic particles with an exponentially growing spectrum of (mass)
states can have, beyond which the partition function diverges \cite{Hagedorn:1965st}.  
Beyond $T_H$ deconfinement will start, and, depending on the undersaturation of quarks in the matter
\cite{Stoecker:2015zea} a Yang Mills plasma or a 2+1 flavour Quark Gluon Plasma (QGP) will form. 
$T_H$ in the present approach is identified with the critical temperature $T_c$. The Hagedorn states (HS) 
within the SBM are the presently not yet discovered heavy
``missing hadron states'' which can be attributed to the exponential part of the Hagedorn spectrum and 
which are most abundant in the vicinity of $T_H$. 
In
\cite{Greiner:2004vm} HS being created in multiparticle collisions are proposed to serve 
as a tool for a microscopic description of the phase transition from HRG to QGP. 
The HS can have any quantum number combination compatible with their mass.
This property of HS was applied in \cite{NoronhaHostler:2007jf,
NoronhaHostler:2010dc,NoronhaHostler:2009cf}  
in order to understand why (multi-) strange 
(anti-) hyperons at the Relativistic Heavy Ion Collider (RHIC) 
chemically equilibrate much faster than the typical life time of a fireball $\left( \sim10\fmc
\right)$. In the vicinity of $T_H$ the most abundant mesons, i.e. pions and kaons, do 'cluster' to Hagedorn
states which in turn can decay into several kinds of hyperons. Via a coupled set of rate equations, one
for each species, the chemical equilibration times for $p,K,\Lambda$ are of order $t\sim1-2\fmc$. 
The inclusion of Hagedorn spectra in the partition functions of the HRG provides a lowering of the speed of
sound, $c_s$, at the phase transition and a significant decrease of the shear viscosity to entropy
ratio $\eta/s$ \cite{NoronhaHostler:2008ju,NoronhaHostler:2012ug}. This results are in good
agreement with corresponding lattice calculations
\cite{Majumder:2010ik,Jakovac:2013iua,Itakura:2008qv}. 
The general impact of HS on the occurrence of various phases from HRG to 
deconfined partonic matter was also studied in various MIT bag model descriptions 
\cite{Moretto:2005iz,Zakout:2006zj,Zakout:2007nb,Ferroni:2008ej,Bugaev:2008iu,Ivanytskyi:2012yx,Vovchenko:2015cbk}. 
There the Hagedorn spectrum $\rho\left( m \right)$ with  
\begin{equation}
	\rho\left( m \right)=f\left( m \right)\exp\left( \frac{m}{T_H} \right)
	\label{eq:hagspadhoc}
\end{equation}
is applied, where the pre-function $f\left(m  \right)$ mimics the low-mass part 
of the Hagedorn spectrum. 
\newline
\newline
To describe the
hadronization of jets in $e^+$-$e^-$ annihilation events the concept of color strings
\cite{Andersson:1983ia} was applied. Here the basic idea is that partons tend to cluster in color
singlet states from the very beginning of the generated event. These clusters decay to smaller
clusters until some cut-off scale is reached and hadrons are formed
\cite{Webber:1983if,Marchesini:1991ch}. In the framework of RQMD multi-particle collisions and
decays were considered as particle clusters for which separable interactions have to exist
\cite{Sorge:1989dy}. A statistical approach \cite{Werner:1995mx} considers the hadronization of
quark matter droplets to hadrons within the microcanonical ensemble. According to a $n$-body phase 
space these quark matter droplets decay via Markov chains into various hadron configurations. A
further statistical treatment of HS was performed in Ref.~\cite{Pal:2005rb} using a simplistic
description of the Hagedorn spectrum. The authors regarded one
single massive $\left( m\sim100\GeV \right)$ initial resonance which consecutively cascades down 
via decay chains until only stable hadrons as protons, neutrons, pions etc.~are left. 
The various terms like 'clusters',
'quark matter droplets','massive resonances' or 'bags' could be identified with possible Hagedorn
states.  
\newline
\newline
Following our recent approach \cite{Beitel:2014kza} we here implement for the first time Hagedorn states in
microscopical dynamical box simulations: 
In contrast to a non-covariant bootstrap equation \cite{Frautschi:1971ij,Hamer:1971zj} we have employed 
a covariant one
\begin{align}\label{eq:tautotbsq}
  &\tau_{\small{\vec{C}}}\left(m\right)=\frac{R^3}{3\pi
  m}\sum\limits_{\small{\vec{C}_1,\vec{C}_2}}\,\int\limits_{m^0_1}^m
  \dd m_1\mkern-15mu\int\limits_{m^0_2}^{m-m_1}\mkern-15mu\dd m_2\,\tau_{\small{\vec{C}_1}}(m_1)\,m_1\\
  &\times\tau_{\small{\vec{C}_2}}(m_2)\,m_2\
  p_{cm}\left(m,m_1,m_2\right)\delta^{\left( 3 \right)}_{\vec{C},\vec{C}_1+\vec{C}_2}
  ,\nonumber
\end{align}
where
$\tau_{\vec{C}}\left( m \right)$ denotes the mass density of Hagedorn states with charge $\vec{C}=\left( B,S,Q
\right)$ and mass $m$, where $B$ is the baryon number, $S$ the strangeness and $Q$ the electric
charge. The terms $\tau_{\vec{C}_1}$ and $\tau_{\vec{C}_2}$ stand for spectra of both constituents 
which make up spherical HS with radius $R$ whose density is described by $\tau_{\vec{C}}\left( m \right)$. 
In the rest frame of created HS, $p_{cm}$ denotes the momenta of the decay products and ensures
strict energy-momentum conservation. Exact charge conservation is applied too. 
This highly non-linear integral equation of Volterra type is solved numerically. The initial input 
for $\tau_{\vec{C}_{1,2}}$ are spectral functions of the hadronic transport model UrQMD
(Ultrarelativisc Quantum Molecular Dynamics) \cite{Bass:1998ca} consisting of 55 
baryons and 32 mesons. Inserting the hadronic spectral functions into the r.h.s.~of Eq.~\ref{eq:tautotbsq}
results in first HS consisting of two hadrons only. In the subsequent steps, these created HS serve
as constituents of next heavier HS, which now may consist of one HS and one hadron or of two lighter HS.
This means that every Hagedorn spectrum on the l.h.s.~sooner or later will appear as constituent on
the r.h.s.~of Eq.~\ref{eq:tautotbsq} to create next heavy HS. The upper Hagedorn spectrum mass $m$ is increased by $\Delta
m=0.01\GeV$ until a final value of $m=8.6\GeV$ is reached due to computational limitations. 
Numerical results of
Eq.~\ref{eq:tautotbsq} are provided in \cite{Beitel:2014kza}. We find that the Hagedorn
temperature rises when $R$ gets smaller and is very weakly dependent on charges $\vec{C}$. 
In addition with the Hagedorn spectrum we were able to derive an expression for HS
total decay width $\Gamma$
\begin{align}\label{eq:gamgen}
  &\Gamma_{\vec{C}}\left(m\right)=\frac{\sigma}{2\pi^2\tau_{\vec{C}}\left(m\right)}
  \sum\limits_{\vec{C}_1,\vec{C}_2}\int\limits_{m^0_1}^m\!\dd
  m_1\mkern-15mu\int\limits_{m^0_2}^{m-m_1}\mkern-15mu
  \!\dd m_2\tau_{\vec{C}_1}\left(m_1\right)\tau_{\vec{C}_2}\left(m_2\right)\nonumber\\
  &\times
  p_{cm}\left(m,m_1,m_2\right)^2
  \delta^{\left( 3 \right)}_{\vec{C},\vec{C}_1+\vec{C}_2}.
\end{align}
To compute $\Gamma_{\vec{C}}$ we
applied the principle of detailed balance between binary collisions to create HS and their decays into two
particles, i.~e.~ $2\rightarrow1$ and $1\rightarrow2$ only.  
The limitation to $2\leftrightarrow 1$ processes is necessary 
when implementing HS
into a customary cascade-type transport model which is based on a geometrical interpretation of 
cross sections as it is now implemented in UrQMD. 
 The total decay width is expressed in terms of HS creation cross
section $\sigma$ which are in the range $\Gamma\approx0.4-3.5\GeV$. Various results can be found in \cite{Beitel:2014kza}.
There are some light HS whose total decay width exceeds the mass. 
Their total yield 
is less than 15\%, so we decided to ignore this small effect.
\newline
\newline
The HS with mass $m$, quantum numbers $\vec{C}$, total decay width $\Gamma_{\vec{C}}$ and the
various
branching ratios $\mathcal{BR}$ have , in the present work, been implemented fully into the UrQMD
model. 
The evolution from nonequilibrated initial HS gas through detailed balance between HS creations and
HS decays to equilibrated HRG is simulated in a $10*10*10\fmt$ cubic box with reflecting walls. 
Each simulation is done at an energy density between $\epsilon=0.3-2.0\GeVfmt$ in steps of
$\Delta\epsilon=0.1\GeVfmt$. The quantum numbers of all initial heavy HS in the box are 
assumed to have $\vec{C}=\left( 0,0,0\right)$ to simulate an uncharged gas. The time evolution will therefore
produce all charges $\vec{C}$ only by the decays of HS into charged hadrons and lighter HS, as is
the case in ultrarelativistic heavy ion collisions at RHIC and at the LHC, where all the net charges
at midrapidity are close to zero, e.~g.~ a net baryon density of $\rho_B\approx0$ has been measured. 
Thus the initial ensemble of (heavy) HS creates dynamically all kinds of (light) hadrons until
chemical equilibrium between HS and hadrons is achieved. 
As a 
more conventional alternate conceptual point of view consider the following picture of hadronization 
: In an (ultra-) relativistic 
collision of two heavy ions a large QGP drop is being created. This drop expands, cools 
and decays into smaller droplets/HS close to the transition temperature $T_H\approx T_c.$ 
The HS
propagate, collide with each other and with hadrons, until they decay. 
Among the decay products of the HS at first hadrons will appear quasi instantly. 
The hadrons and the HS may create new HS or hadrons, which then can go on to collide elastically or
inelastically with each other. 
As long as the system stays at a high temperature $T\approx
T_H$ the dynamical interplay between hadrons and HS will drive both into thermal and chemical
equilibrium. To examine this process we distribute initial HS uniformly in momentum and
configuration space and demand: 
\begin{equation}
	E=\sum\limits_{i=1}^{N_0}E_i,\quad\vec{p}=\sum\limits_{i=1}^{N_0}\vec{p}_i=0
	\label{eq:consvlaw}
\end{equation}
The initial number of HS is set to
$N_0=200$. Each run lasts
until $t=20\fmc$ and all results are averaged over 1200 runs. 
\newline
 Fig.~\ref{fig:hsmassdistprl} shows the
final mass distribution of all HS which arises when a system with exponential growth of mass states is
subjected to the Boltzmann distribution. The equilibrated mass distribution of HS shown in 
Fig.~\ref{fig:hsmassdistprl} results from the 
convolution of the Hagedorn spectrum $\tau\sim\exp\left( m/T_H \right)$ and the 
Boltzmann distribution $f\left( E\right)\sim\exp\left( -E/T \right)$. Observe that for higher 
energy densities more HS with higher masses are formed. 
The impact of HS on the system's total particle number, energy and mass after $t=5\fmc$ is shown in 
Fig.~\ref{fig:tempeps}. At the largest energy density, $\epsilon=2.0\GeVfmt$, already every fourth
particle is a HS and more than $\sim60\%$ of the total energy and $\sim70\%$ of the total mass are occur
by HS. All results are in full accordance to the SBM: HS appear most abundantly at
$T_H$, which is reached when $\epsilon\rightarrow\infty$. We also observe that with increasing
energy density more and more of the available total energy is converted into (massive) HS rather than being
distributed over the kinetic degrees of freedom.
The latter result is backed by the dependence of hadrons' Boltzmann slopes $T$ on $\epsilon$ as shown
in the lower part of Fig.~\ref{fig:tempeps}. 
Increasing energy density $\epsilon$ causes the temperature $T$ of all hadrons to converge to the
Hagedorn temperature $T_H$. This result contradicts the usual HRG thermodynamics, where the
temperature grows with the energy density beyond any limit. Our result manifests the SBM statement
\begin{equation}
	\lim\limits_{\epsilon\to\infty}T=T_H.
	\label{eq:temlim}
\end{equation}
Note that decay chains of a single massive HS already show slopes with a Hagedorn temperature
\cite{Beitel:2014kza}, - a consistent picture. 
\begin{figure}
\begin{tabular}{c}
\includegraphics[angle=270,width=0.5\textwidth]{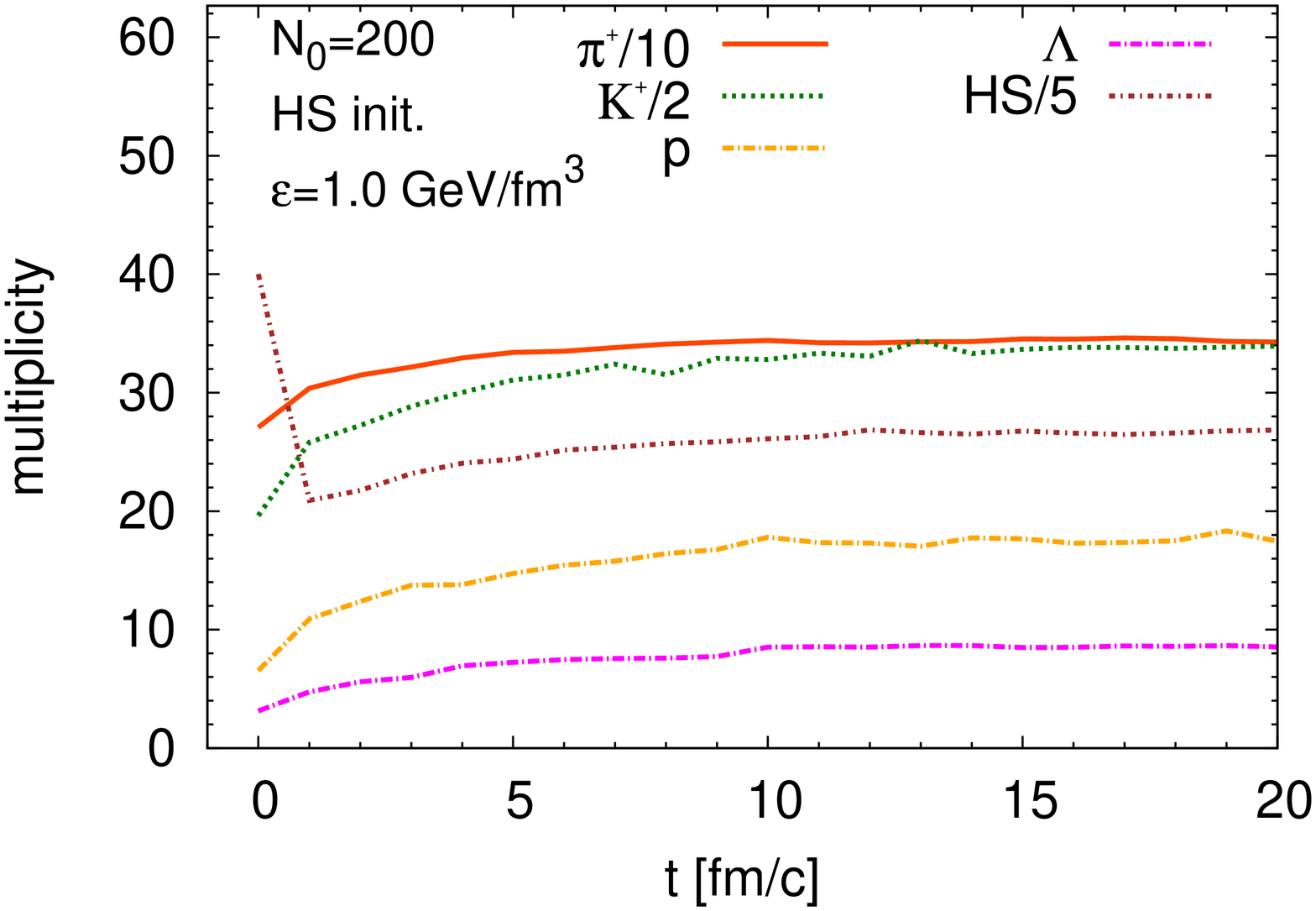}\\
\includegraphics[angle=270,width=0.5\textwidth]{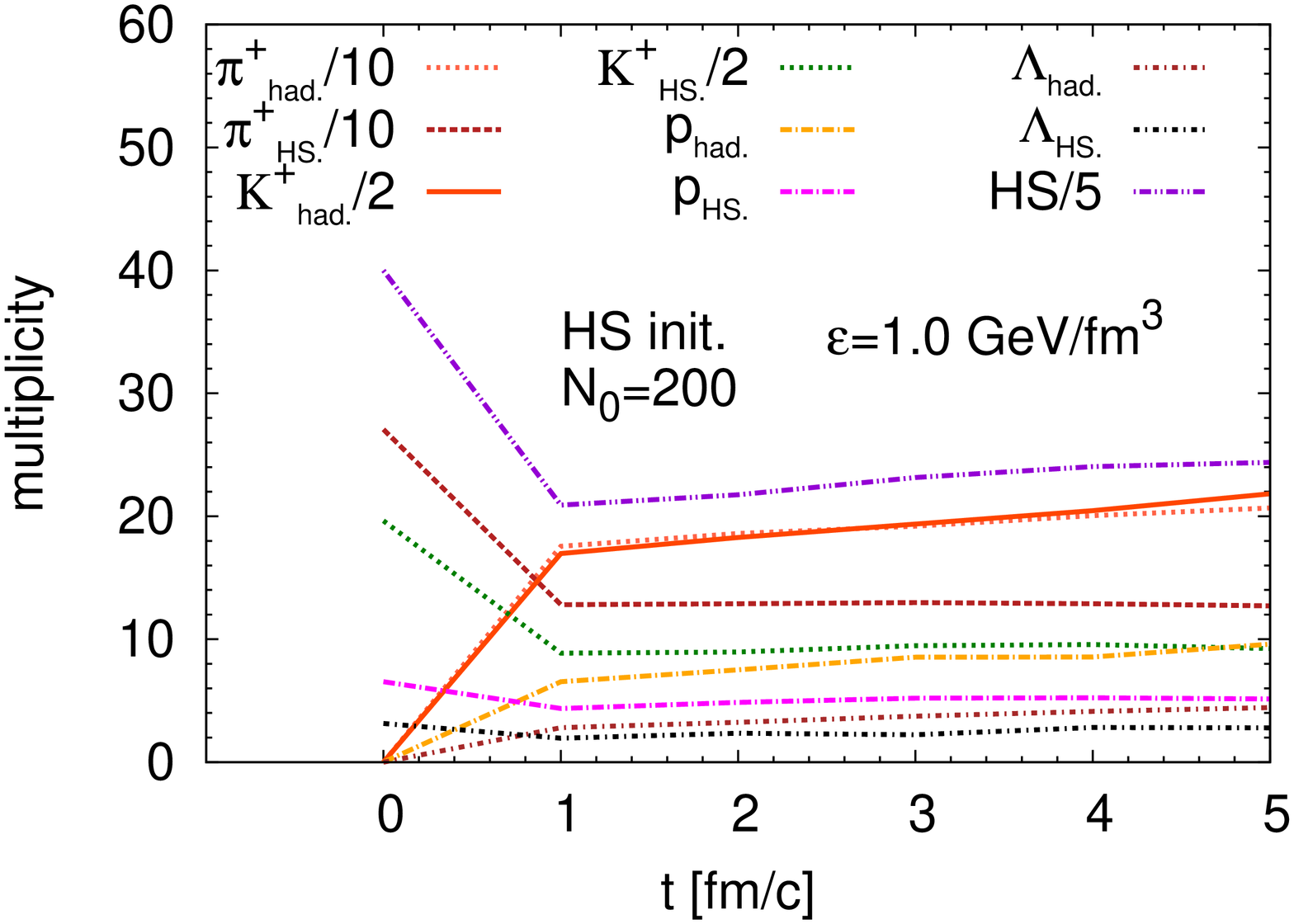}
\end{tabular}
\caption{Time evolution of total (feed down corrected)  multiplicities (top) of $\pi^+$, $K^+$, $p$  and
$\Lambda^0$ at energy density $\epsilon=1.0\GeVfmt$. Time evolution (bottom) of first 5 \fmc of direct multiplicities plus
feed down contributions from resonance decay (had.) and from HS decays (HS.) for same hadrons as
mentioned above. In case of HS
direct multiplicities were considered.}
\label{fig:mulprl}
\end{figure}

\begin{figure}
\begin{tabular}{c}
\includegraphics[angle=270,width=0.5\textwidth]{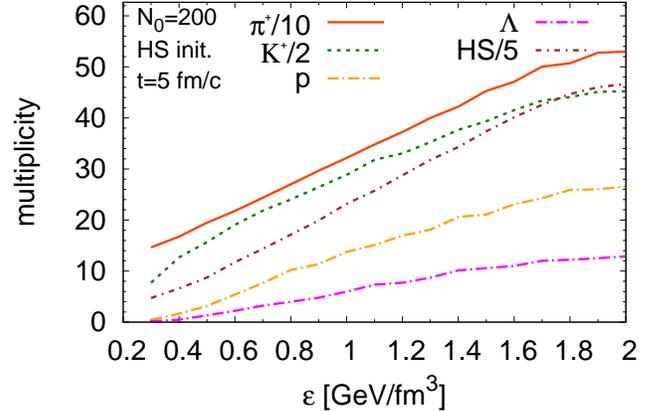}
\end{tabular}
\caption{Total multiplicity dependence on energy density
 of
$\pi^+$, $K^+$, $p$ and $\Lambda^0$ close to chemical equilibrium $\left(t=5\fmc\right)$. In case of HS
dynamical multiplicities were considered.}
\label{fig:mulprl2}
\end{figure}
\begin{figure}
  \includegraphics[angle=270,width=0.50\textwidth]{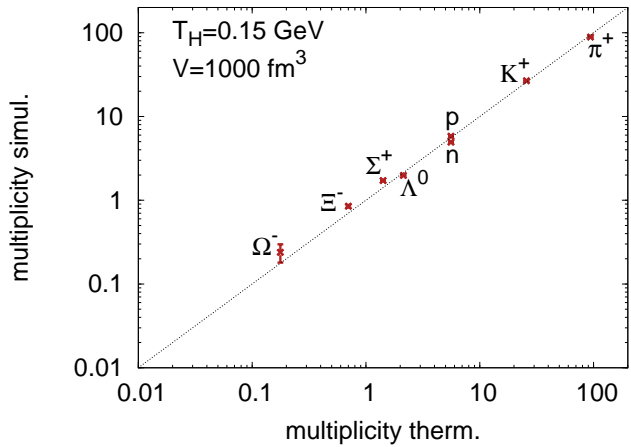}
\caption{Hadronic multiplicities from dynamical box simulation (ordinate) for
$\epsilon=1.0\GeVfmt$ $\left( T\approx0.154\GeV \right)$ at $t=20\fmc$ and corresponding thermal
model (abscissa) for $T=0.154\GeV$ and $V=1000\fmt$. Statistical error in simulations is 
for $\Omega^-$ roughly $25\%$ and for others less than $5\%$.
}
\label{fig:thermcompprl}
\end{figure}

Observe the very fast thermalization time 
$t=1-2\fmc$, where hadrons, hadron resonances and HS interact rapidly, changing energy and quantum
numbers.
Fig.~\ref{fig:mulprl} shows the time evolution of pions, kaons, protons and lambdas
as 'direct' decay hadrons and feed down of hadron resonances and of HS. The later feed down 
corresponds to (potential) hadronic particles 'stored' in the existing HS as calculated via their decay
chains as discussed in \cite{Beitel:2014kza}. 
The very fast thermalization in our simulations 
of $t\le2\fmc$ is now obtained from the initial decaying HS in the system. The upper figure shows
that the number of HS drops down
within $t=1\fmc$ and then saturates. The emerging hadrons and hadronic
resonances are build up during by these decays and by the subsequent regenerations of HS on such
short time scale (lower figure). 
The sum of the yield hadrons stemming from feed down of the HS and of hadronic resonances (shown in
the lower figure) accounts for the total stable particle yields in the upper figure. Within
$t=1\fmc$ more than a half of the initial HS has decayed into hadrons reaching a stationary value at
$t=1-2\fmc$ and a further moderate saturation. The very fast chemical equilibration occurs 
by means of decays, recreation and rescatterings of
HS in such a dense environment. This is in contrast to standard hadronic transport approaches of 
\cite{Belkacem:1998gy,Bratkovskaya:2000eu}.
\newline
Fig. \ref{fig:mulprl2} shows that in chemical equilibrium total multiplicities rise nearly
linear with energy density in the evolving system originating from the initial assembly of HS. 
The yields are determined predominantely by the particle's masses. 
HS exhibit the steepest slope as demanded by SBM.  
Tab.~\ref{tab:ratios} confronts the calculated hadron multiplicity 
ratios at different energy densities with corresponding experimental values as obtained by the 
ALICE collaboration at the LHC. 
\begin{table}[!htbp]
\begin{tabular}{lcccccc}
	\hline\hline
	& p-p & Pb-Pb & 0.3 & 0.8 & 1.0 & 2.0 \\
	\hline
	$K^-$/$\pi^-$ & 0.123(14) & 0.149(16) & 0.192  & 0.197 & 0.193 & 0.185\\
	$\bar{p}$/$\pi^-$ & 0.053(6) & 0.045(5) & 0.015  & 0.049 & 0.052 & 0.060\\
	$\Lambda$/$\pi^-$ & 0.032(4) & 0.036(5) & 0.007  & 0.022 & 0.024 & 0.029\\
	$\Lambda$/$\bar{p}$ & 0.608(88) & 0.78(12) & 0.475  & 0.456 & 0.469 & 0.499\\
	$\Xi^-$/$\pi^-\!\!*10^{3}$ & 3.000(1) & 5.000(6) & 1.565  & 6.492 & 5.769 & 7.106 \\
	$\Omega^-$/$\pi^-\!\!*10^{3}$ & - & 0.87(17) & 0.137  & 0.815 & 0.823 & 0.994\\
	\hline\hline
\end{tabular}
\caption{Comparison
  of particle multiplicity ratios from theory vs.~p-p at $\sqrt{s_{NN}}=0.9$\TeV \cite{Aamodt:2011zza} 
  and Pb-Pb at $\sqrt{s_{NN}}=2.76$\TeV \cite{Abelev:2013vea,Abelev:2013xaa,Abelev:2013zaa}, both from 
  ALICE at LHC. Calculated values are listed for some energy densities in the range 
  $\epsilon=0.3-2.0\GeVfmt$. 
  Numbers in brackets
  denote the error in the last digits of the experimental multiplicity ratios. The 
  statistical error is less than 25\% for strange baryons.}
\label{tab:ratios} 
\end{table}
To demonstrate that a real thermal (chemical+kinetic) equilibrium is reached we compare in 
Fig.~\ref{fig:thermcompprl} simulated hadron multiplicities at $\epsilon=1.0\GeVfmt$ with multiplicities 
provided by a standard HRG thermal model. A temperature of $T\sim0.154\GeV$ is assumed in the thermal
model and for
$\epsilon=1.0\GeVfmt$ in the full evolution as depicted in Fig.~\ref{fig:tempeps}.  
The perfect agreement supports that our results thermally equilibrated. 
\newline
\newline
In summary, it was shown that a system of HS, e.~g.~ as emerging of large QGP drops
created in heavy ion collisions gives a new insight how hadronization can take place. Starting in
non-equilibrium dynamical decay and (re-) creation of HS provide on a very short time scale of
$t=1-2\fmc$ all hadrons of the HRG as advocated
over the years in \cite{BraunMunzinger:2001ip,Andronic:2011yq}. Potential decays of HS might also explain the
finding of $e^--e^+$ \cite{Becattini:1995if} and $p-\bar{p}$ \cite{Becattini:1997rv} within a
thermal model analysis. The implementation in microscopic transport models of full heavy ion collisions
sets a new venue at future investigations, also for finite net baryon densities at NICA facilities
in Dubna and CBM experiments at the FAIR facility which is build adjecent to the GSI in Darmstadt. 
Multibaryonic and multistrange HS can serve as an energy
reservoir for production of rare hadronic particles. The implications of HS on the shear viscosity
and on the net baryon number fluctuations in dense hadronic matter has to be  studied in the future. 
\newline
\newline
We thank K.~Gallmeister for critical discussions. 
This work was supported by the Bundesministerium f\"ur Bildung und
Forschung (BMBF), the HGS-HIRe for FAIR and GSI, the Helmholtz International Center
for FAIR within the framework of the LOEWE program launched by the
State of Hesse. Numerical computations have been performed at the
Center for Scientific Computing (CSC) at the Goethe Universit\"at Frankfurt.
\bibliography{refs}

\end{document}